# The ET INTERVIEW: PROFESSOR JOEL L. HOROWITZ

*Interviewed by Sokbae Lee*
*Columbia University*

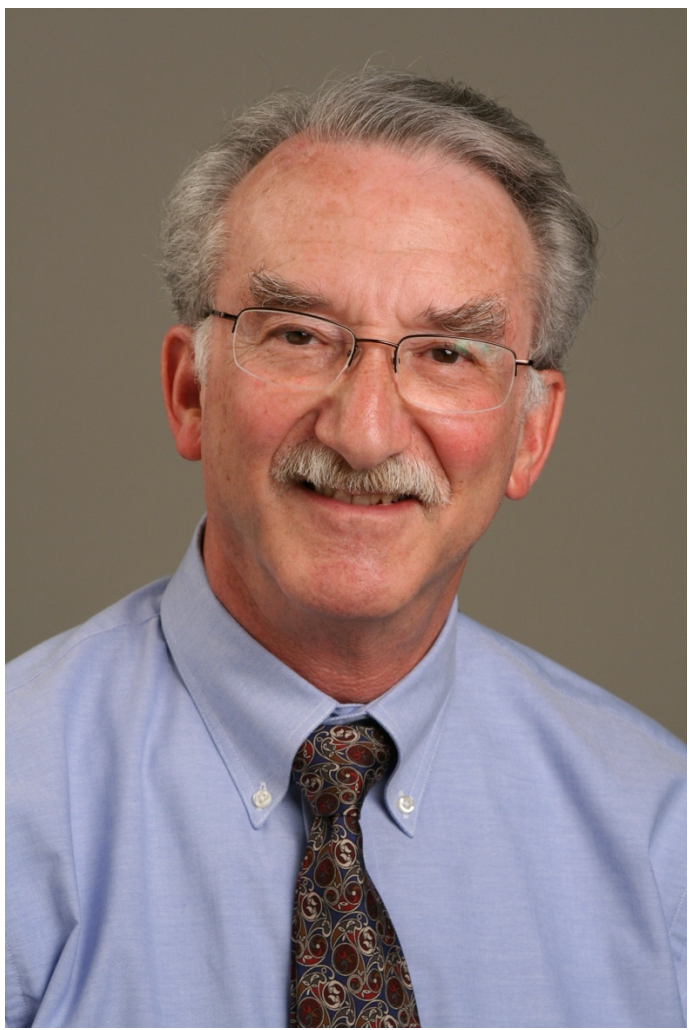

(Photo caption: Joel L. Horowitz)

*Joel L. Horowitz has made profound contributions to many areas in econometrics and statistics. These include bootstrap methods, semiparametric and nonparametric estimation, specification testing, nonparametric instrumental variables estimation, high-dimensional models, functional data analysis, and shape restrictions, among others. Originally trained as a physicist, Joel made a pivotal transition to econometrics, greatly benefiting our profession. Throughout his career, he has collaborated extensively with a*

*diverse range of coauthors, including students, departmental colleagues, and scholars from around the globe.*

*Joel was born in 1941 in Pasadena, California. He attended Stanford for his undergraduate studies and obtained his Ph.D. in physics from Cornell in 1967. He has been Charles E. and Emma H. Morrison Professor of Economics at Northwestern University since 2001. Prior to that, he was a faculty member at the University of Iowa (1982-2001). He has served as a co-editor of Econometric Theory (1992-2000) and Econometrica (2000-2004). He is a Fellow of the Econometric Society and of the American Statistical Association, and an elected member of the International Statistical Institute.*

*The majority of this interview took place in London during June 2022.*

**Early Life and Education**

*Let's start with your early life and your family background.*

My father, Norman, was a professor of biology, specifically a geneticist. He got his PhD at Caltech in 1939. After receiving his Ph.D., he became a postdoc with a geneticist named George Beadle, who was a professor of biology at Stanford. Our family moved to Palo Alto, and my father spent World War II working in Beadle's laboratory at Stanford. I remember my father helping Beadle in his victory garden, which he grew to reduce demand for other sources of vegetables. I was too young to help in the garden. One of the things Beadle's lab worked on was making penicillin. That kept them all out of the army. Beadle had been at Caltech in the 1930s and moved back there after the war. My father went with him and became an assistant professor of biology there. He spent his entire career at Caltech. At that time, many senior people in biology did not believe in the existence of genes. This was before Watson and Crick and, of course, before molecular biology. Researchers had to do very clever, wet chemical experiments with organisms like corn, bread mold, fruit flies, all of which reproduce relatively quickly. My father was one of the first people (maybe the first) to do an experiment that confirmed a hypothesis that Beadle, in collaboration with Edward Tatum and Boris Ephrussi, had proposed. It is called the one gene-one enzyme hypothesis: each gene codes for one enzyme. The experiment was one of the first to establish the existence of genes. It was courageous for a junior biologist to do this when there were senior people in the field who did not believe in the existence of genes. Of course, after Watson and Crick, everything changed. My father did genetics research until the early 1960s, when he became interested in the possibility of life on Mars. Because of that interest, he became the head of the Space Biology Division at the Caltech Jet Propulsion Laboratory. He divided his time between Caltech and JPL. During his time at JPL, he developed a life detection experiment that flew on the Viking lander missions to Mars.

*Is this after you were born?*

Oh, yes; this was in the mid-1970s. I had finished my Ph.D. and was living in Arlington, VA, which is a suburb of Washington DC.

*Just circling back, when you were born, where was your father?*

He was still in Pasadena. Then the family moved to Palo Alto. My sister, Elizabeth, was born there. We moved back to Pasadena in 1946. The one gene-one enzyme experiment happened in 1944. My father had long been interested in the origins of life, and this led him to be interested in the possibility of life on Mars. His work at JPL began in 1965. It was very controversial. There were several different life detection experiments on Viking, and the investigator responsible for one claimed to have detected life. Few people believed that life had been detected, and the official NASA record of the Viking missions says that they did not detect life. My father eventually went back to Caltech full time and became the head of the biology department. In the meantime, his whole field had changed. Molecular biology had taken over genetics while he was at JPL doing other things. He did not have the technical skills needed to do experiments in molecular biology. In addition, my mother had had a severe stroke. So, he retired to take care of her, though he kept an office at Caltech and attended seminars.

*Can you also tell us a little bit about your mother?*

My mother had a master's degree in biology from Brown University. She grew up in a relatively poor, Orthodox Jewish family in the Dorchester neighborhood of Boston. She had three sisters and was the only one who finished college. She went to high school at Girls' Latin School in Boston and then went to Radcliffe College. At that time, Harvard did not admit women; Ivy League schools had sister women's colleges. Radcliffe was the sister college of Harvard and was close to the Harvard campus. It merged with Harvard in 1999 and is no longer a women's college. After Radcliffe, my mother went to Brown to get a master's degree in biology. She and my father met at Woods Hole, Massachusetts. Woods Hole is on Cape Cod and has an important oceanographic laboratory. In 1938, they had their honeymoon in Pacific Grove, California, which is contiguous with Monterey. She was a homemaker throughout the war. She turned to part time work in the 1950s. At first, she was a research assistant to a member of the engineering faculty at Caltech. Then she became a social worker for Los Angeles County. But as a child, she had rheumatic fever, which damaged her heart valves. Due to later complications of that, she had a severe stroke in her early 60s that left her unable to talk or walk.

*And you have one sister?*

Yes, I have one sister, Elizabeth. She is three and a half years younger than I am. She graduated from UC Berkeley and has lived in Berkeley for all her adult life. She went to nursing school after our mother's stroke and was a nurse at Children's Hospital Oakland for many years. She retired several years ago and still lives in Berkeley.

*Shall we now briefly talk about your student life before college?*

I went to Pasadena High School, which at that time was on the campus of Pasadena City College. We were treated like college students and had a lot more freedom to move around the campus than high school students have today. My friends were what you might call nerdy kids. We took all the honors courses. We had a course in solid geometry but after about two days, it seemed to us that solid geometry was a useless subject. So, we persuaded the teacher to teach us calculus instead.

*Which subject was your favorite in high school?*

Math and physics were probably my favorite subjects.

*I guess that continues into college. Did you major in physics at Stanford?*

Yes. Majoring in physics gives you a view of things that is different from the view in economics. In physics, a theory means something that predicts the results of experiments that haven't yet been done. Every theory is a target. In economics, by contrast, utility theory is sort of kept under glass and is very flexible. If an experiment gives results inconsistent with utility theory, many economists look for ways to rationalize the results or treat them as anomalies. But in physics, theories are targets. When I was a graduate student, the only theory, meaning something that predicted the outcomes of experiments that had not yet been done, was quantum electrodynamics. Quantum electrodynamics is a theory of the interaction of light and matter. It made predictions with an accuracy of something like 10 significant figures. Every experimentalist knew that if he (they were all men at that time) could find an error in quantum electrodynamics (not a mathematical error but a prediction error) he would win a Nobel Prize. So people designed instruments that could detect deviations from predictions of the order of a few times ten to the minus ten.

*It's a very precise statement!*

The theory worked that well, but if you could find something wrong with it – bang it would be gone! Nobody found a deviation from prediction. While I was a graduate student, Val Fitch from the University of Illinois gave a seminar about an experiment he and his collaborators had done that violated time reversal invariance. Time reversal invariance was a fundamental principle of theoretical physics. People asked a lot of questions along the lines of "Did you do the experiment correctly?" and "Was there a measurement error?" After he persuaded people that the experiment was correct, the theorists got excited. They said, "Here's a new thing for us to do. Let's figure out what's going on here and find a new theory that takes this into account." Fitch and his collaborator, James Cronin, won a Nobel Prize. It's a very different attitude toward theory.

*So experimental results are confirmed as true findings, and then the theory people move in to explain.*

This also happened in 1956, when Chien-Shiung Wu carried out an experiment proposed by the theoretical physicists Tsung-Dao Lee and Chen-Ning Yang in which conservation of parity was violated. The existing theory said, roughly speaking, that there is no difference between the behavior of the actual world and its mirror image. But the experiment showed that there is a difference. People had to find out whether the experiment had been designed correctly and so forth, but after that had been done, theory was modified to accommodate non conservation of parity. Lee and Yang won the Nobel Prize in physics for this discovery.

*That is totally different from economic theory, I guess. Right?*

Yes. Economic theory is, at least in part, a substitute for precise measurement and "laws" like the laws of physics that are universally valid. There is beginning to be more respect in economics for experimental results, due in part to improved understanding of identification issues that are involved in economics and the results of experiments in development economics. Joshua Angrist, David Card, and Guido Imbens won the Nobel Prize in economics for their work showing how some of the identification problems can be overcome and the kinds of empirical results that can be obtained.

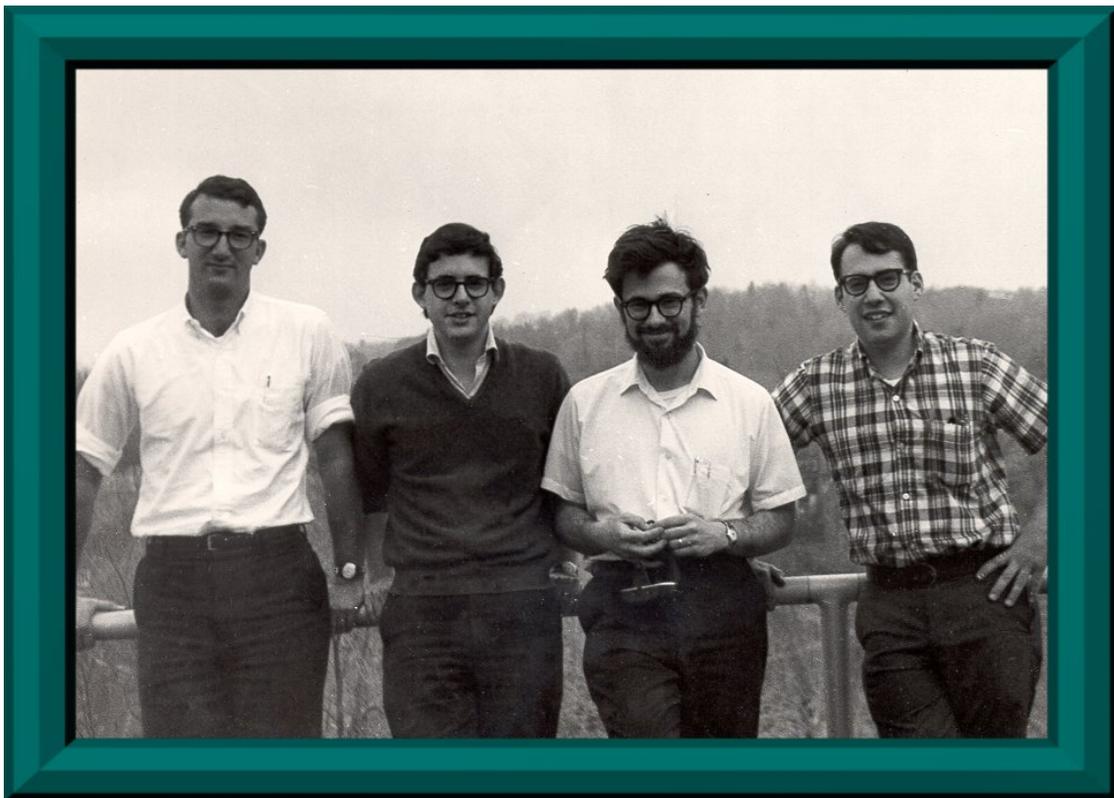

(Photo caption: Joel (farthest to the left) with other graduate students on the balcony of the Laboratory of Nuclear Studies at Cornell)

*Coming back to your PhD work: maybe you can elaborate a bit about your thesis and your advisor?*

My advisor was a man named Kenneth Wilson. He got his Ph.D. at Caltech in 1961, moved to Cornell in 1963, and became a full professor in 1970. That was 3 years after I had finished my Ph.D. there. He was very young. We used to go folk dancing together. One day he went off to the chemistry department and stayed there for a year or two. He came back with ideas about critical phenomena and phase transitions that won him the Nobel Prize in physics.

*Was it during your PhD period?*

He won the prize in 1982, long after I had finished my PhD. The prize was for work he had done while I was at Cornell but not in my presence because he spent much of his time in the chemistry department.

*So, your thesis was about something else?*

Yes. My thesis was about a problem in high energy theory concerning interactions between mesons.

*Was it mainly involved with theory like mathematics or with experiments?*

It was about theory and was mathematical, but mathematics in physics is treated differently from mathematics in economics. The level of mathematical rigor in economics is much greater than in physics. For example, quantum electrodynamics is about the interaction of light and matter, but mathematically, it is a system of four simultaneous partial differential equations in four dimensions: three spatial dimensions and time. The dependent variables in these equations are not functions of the usual kind. They are operators in a space that would be a Hilbert space, except it contains delta functions. Nobody knew when I was a graduate student and I don't think anybody knows to this day if these equations have a solution. But there is a way to solve differential equations by turning them into integral equations. In quantum electrodynamics the integral equations contain a dimensionless constant called the fine structure constant whose value is roughly 1/137. The solution method consists of making a series expansion in powers of the fine structure constant. The first few terms of this expansion give the results of classical electrodynamics (e.g., Maxwell's equations) and the famous experiments of the early 20th century in nuclear physics and quantum theory (e.g., pair production and Compton scattering). However, the higher order terms are all infinite. Richard Feynman, Julian Schwinger, and Shinichiro Tomonaga independently developed a method called renormalization that amounts to subtracting infinity from infinity in a way that yields 10 significant figures of accuracy. Feynman, Schwinger, and Tomonaga shared the 1965 Nobel Prize in physics for this work. Renormalizability became a necessary condition for acceptability of a quantum field theory. When I was a graduate student, nobody cared whether there was a solution to the equations of quantum electrodynamics or a mathematically rigorous theory of renormalization. Renormalization was consistent with intuition about how the relevant interactions worked, was mathematically correct to the extent that was possible to discern at the time, and was accurate to 10 significant figures.

*It seems like it is quite difficult - very high-level math.*

It is high level but when the mathematics become intractable, it is OK to rely on intuition that is consistent with the results of experiments. When I was a student, I wanted to find mathematically rigorous foundations of quantum mechanics. I went to the math department and took a course in functional analysis, which seemed like the right thing to do for this topic. When I told the theoretical physics faculty that I wanted to make quantum mechanics mathematically rigorous, they asked why I wanted to do that. They said that if you have a theory or procedure that gives 10 significant figures of accuracy and an intuitive interpretation of it, you have a whole story. Several years ago, a string theorist told me that it was still unknown whether the equations of quantum electrodynamics have a solution. This was some 50 years after I finished my Ph.D. I doubt that anybody really cared if the equations have a solution.

*Because they can do whatever they want without relying on this fundamental existence theorem?*

Yes.

*That's kind of an interesting comparison. Your PhD work is mainly theoretical but quite a different type of math. But still, I guess you took functional analysis and this is highly related to your future work as an econometrician.*

Yes. Toward the end of my graduate career, I decided I wanted to do something that was closer to the real world and did not require spending millions of dollars on complicated machines and instruments to understand the phenomena I was studying. So after my Ph.D. I worked first at a defense consulting company and later at the U.S. Environmental Protection Agency for about 15 years. But these days, I say I'm doing many of the things I did as a theoretical physicist, except the names of the variables are different.

*That's quite an interesting way of looking at your whole career. Do you have anything you want to add about the period before your Ph.D.?*

I will add one thing. Undergraduate training in physics, at least, at Stanford, was quite different from the way undergraduate training in economics works. It was more like training in engineering now. Stanford was on the quarter system. As a first quarter freshman majoring in physics, you took a calculus course but not a physics course. In the second quarter, you took the first physics course and the second calculus course. The physics course used the calculus that you had just learned in the first quarter. It continued that way. In one quarter you took a mathematics course, and it was used in the next quarter's physics course. In the second year, you took courses in ordinary and partial differential equations. The third-year physics courses were quantum mechanics and classical electricity and magnetism. These made extensive use of differential equations. If you hadn't taken the mathematics courses, you had no idea what was going on in the physics courses and would have to drop out of the physics major, because the fourth-year physics courses assumed you had taken and understood the third-year courses.  You couldn't decide in, say, your third year to change your major from English or biology, for example, to physics, because you were supposed to know the mathematics and physics taught in the first two years by the time you reached your third year. It is possible to change your major to economics from another subject in many departments. This makes graduate training in economics very different from undergraduate training and may account for the shock some new graduate students in economics experience when they find out what is involved in graduate training in economics.

Another difference is that in our junior year we had 6 eight o'clock courses - eight o'clock in the morning, Monday through Saturday.

*Even on Saturday?*

Yes, throughout the entire academic year.

*It must have been a totally different university at the time.*

Saturday classes were not unusual.

*People used to work more. Did you still have time to gather and have some parties?*

Yes, but you didn't want to stay up until two or three in the morning on a Friday night.

*I guess you needed to show up the next morning.*

If you didn't show up, you could be in trouble for having missed something. They did not take roll, but the course moved quickly. If you missed something, you would be lost for the rest of the course. Another memorable feature of the freshman physics classes, not the more advanced classes, was the demonstrations. One course spent several weeks on optics. One day there was a concave mirror on the laboratory bench with a lit light bulb in a socket in front of it.  A student arrived late and the professor asked him to turn off the light by unscrewing the bulb.  The student reached over to do that but there was no light bulb. The bulb was at the back of the lecture hall. The image that the student reached for was at the focal point of the mirror. That was an impressive illustration of an important property of concave mirrors.

*I see! So that was like magic. That was a demonstration of optic power!*

The professor who taught a freshman course in classical mechanics wanted to demonstrate in a convincing way that the amplitude of an oscillating pendulum doesn't change. There was a pendulum with a huge steel bob suspended from the ceiling. The professor, who was the head of the physics department, put his head and body against the blackboard so he could not move farther back. Then he put the pendulum bob against his nose and let it go. The bob swung out and back. It touched his nose on the backswing but didn't bruise or break it. It was a very effective demonstration. Of course, the professor had to be careful not to give the pendulum a little push when he let it go.

*That was a very nice way of demonstrating this principle of physics.*

In the first quarter of my senior year, I took a physics course from a professor named Robert Hofstadter. He won the Nobel Prize in physics during the course. (He shared the prize with Rudolf Mössbauer.) Late in the course, he flew to Stockholm for the prize ceremony. Everyone in the class applauded him when he returned. He told us that he had been accommodated in a fancy hotel in Stockholm. He went to sleep early on the night of his arrival (jet lag) but was awoken around midnight by young women in white gowns with candles in their hair marching around his room. It took some time for him to understand what was happening. It was St Lucia's day. She is the traditional bearer of light in dark Swedish winters.

**Early Career and Transition to Econometrics**

*What was your first job after your PhD?*

After I got my PhD I did not want to get an academic job in physics. I got a job with a defense consulting company called Research Analysis Corporation (RAC). It was located in McLean, Virginia, which is a suburb of Washington, D.C. RAC, which no longer exists, was the Army's version of the RAND Corporation. After World War II, the military realized that it could not recruit the best people to be members of the civil service or the military because there were too many restrictions and salaries were too low. So the RAND Corporation was formed to work for the Air Force, the Center for Naval Analyses to work for the Navy, RAC (originally the Operations Research Office of Johns Hopkins University) to work for the Army, and the Institute for Defense Analyses to work for the joint chiefs of staff. I went to RAC and was put in a department that mainly did research in nonlinear programming and queueing theory. Peter Fishburn was also in the department. He worked on decision theory. Two people in the department, Anthony Fiacco and Garth McCormick, won a big prize in operations research, the Lanchester prize, for an algorithm to solve nonlinear programs called the sequential unconstrained minimization technique (SUMT). The army supported this research because its results were used to study the targeting of nuclear weapons and strategic nuclear exchange. You needed several high-level security clearances to do this work. The relation between the calculus of variations and nonlinear programming is an interesting side light of my time at RAC. In physics, I had used the calculus of variations, which treats constrained optimization in a function space. Nonlinear programming is constrained optimization in a discrete space but these two cultures didn't know about each other. I thought this culture gap was interesting.

I lived in Falls Church, Virginia, at that time and met my wife, Sue, at an amateur theatre company in Washington. She did some acting and backstage work. I worked on lights. The peak of our theatrical careers was working on "The Sign in Sydney Brustein's Window" at a major Washington theatre.  However, working in the theatre required staying up until after midnight several days a week. I couldn't keep doing that and be effective at my day job, so I gave up theatre.

Companies like RAC were called Federal Contract Research Centers (FCRCs) and had guaranteed yearly contracts that were negotiated with the branches of the military they worked for. There was not competitive bidding for these contracts. By the early 1970s, many private corporations had started doing the same kinds of consulting for the military that the FCRCs did.  The private corporations received contracts through competitive bidding. The Nixon administration decided that the FCRCs' receipt of non-competitive contracts was unfair to private industry and that FCRCs should bid for work competitively. RAC was not good at that, and many people left. The leading members of the research group I was part of moved to the department of operations research at George Washington University. While this was going on, I obtained a contract to work on air pollution problems with the predecessor of the Environmental Protection Agency. The EPA was formed in 1970. I left RAC and moved to the EPA in 1971.

*So you switched to a different job.*

Yes. I was in the part of the EPA that was concerned with air pollution. The person I worked for was a deputy assistant administrator who had a PhD in economics. He eventually became an assistant administrator, which is a position that requires confirmation by the Senate. Watergate happened soon after I started working at the EPA. At that time, automobile emissions were much higher than they are now. The Clean Air Act required the administrator of the EPA to hold public hearings to determine the technological feasibility of greatly strengthening automobile emissions standards and, ultimately, to decide whether to strengthen them. The American automobile industry did not want the standards to be strengthened and claimed that it was not technologically feasible to do so. The Nixon administration supported the industry. The Japanese automobile industry, which was only beginning to have a presence in the US market, said that they could meet the more stringent standards. The South African platinum and palladium industry also supported more stringent standards. Platinum and palladium are used in catalytic converters. The White House opposed more stringent standards, but Watergate so weakened the White House that the EPA administrator, William Ruckelshaus, could decide independently whether to strengthen the standards. In the end, he strengthened them. This led the automobile industry to make many technical innovations, including three-way catalytic converters for all new cars and computer control of mixture and ignition. Another emissions control objective was getting people to drive less and use public transportation more. That became my main occupation at the EPA.

*So that's part of transportation systems analysis, I guess.*

Yes. I got interested in what I learned was called travel demand modeling. I didn't know anything about it, so I asked some people at the Department of Transportation (DOT) to tell me what travel demand modeling was and how it was done. The DOT was only a few blocks from my office. I was met by about 20 people from the Federal Highway Administration, the Urban Mass Transportation Administration, and the Office of the Secretary. They made a presentation that, I think, was intended to persuade me that travel demand modeling was something they did very well and was too complicated for me to understand. They also gave me a large stack of reports from consulting firms that they had contracts with. I took the reports back to my office.

*Without explaining, they just gave them to you.*

Yes, they just gave them to me. I started reading the reports and discovered that they were mostly about gravity models and regressions of mode shares or ridership levels on variables that were either aggregates or not sensitive to policy interventions. I didn't know anything about transportation, but I knew a lot about mathematical modeling and Newton's law of gravitation and that gravity models of travel behavior were nonsense. About two thirds of the way through the stack I came to a report by a consulting firm in Boston called Charles River Associates (CRA). CRA had a policy of not putting the authors' names on its reports, but I learned much later that this report was the manuscript of Domencich's and McFadden's book, *Urban Travel Demand: A Behavioral Analysis*. The report pointed out that gravity and other standard models didn't work well and explained why. It proposed a different approach that consisted of random utility choice modeling and led to a multinomial logit model of mode choice by individuals. The logit model's parameters were to be estimated by fitting the model to data using the method of maximum likelihood. The report presented the log-likelihood function and explained that maximizing it yielded parameter estimates with good properties. In physics when I was a student, there was little statistics. I didn't know what a likelihood function was or why maximizing it yielded parameter estimates with good properties. However, there was a very good technical bookstore on the campus of George Washington University. I bought all its statistics and econometrics books. Henri Theil's textbook "Principles of Econometrics" was one of them. I knew a lot of math, so I could read the books.

*So it's kind of self-study for statistics and econometrics.*

Yes. After reading these books, I decided that I could do travel demand modeling. There is a professional organization in transportation called the Transportation Research Board. It is part of the National Research Council but is also a professional society for transportation. It has an annual meeting, which was where I met Dan McFadden.

*At the time, did you know that the CRA report was written by Dan?*

No but I read the Domencich-McFadden book and got to know Dan pretty well. I also got to know the president of Charles River Associates.

*Through transportation?*

Yes. At that time, random utility and multinomial logit modeling were new and exciting research areas, especially in transportation. Dan had a large grant to forecast ridership on San Francisco's BART transit system before it had been built. Much research on logit modeling took place during or because of this grant.

*Were you still working for the EPA?*

Yes. I was on the staff of the Assistant Administrator for Air Programs and worked in an office called the Office of Policy Analysis with several other people. The Assistant Administrator had created this office to help him evaluate policy recommendations he received from other air programs people at EPA and from consultants. He said our job was to keep him from walking into an open manhole. He told us that if we did this for him two or three days a week, we could do whatever we wanted the other days. This made it possible for me to write and publish papers.

*I see. So you started your academic career in transportation?*

Yes. I also taught courses in operations research at George Washington University, and I visited MIT for a year while I was at EPA. This was possible because of the Intergovernmental Personnel Act, which enabled Federal employees to take what might be called a sabbatical leave to visit universities and other approved organizations. Dan had moved back to Berkeley by then but I got to know others at MIT, including Jerry Hausman.

*What year was this?*

Academic year 1977-78. The Reagan administration took office in 1981 and was anti-environmental. This included Anne Gorsuch, whom Reagan appointed to be the administrator of the EPA. She was the mother of the current Supreme Court Justice, Neil Gorsuch.

In addition, my interests were diverging from the EPA's mission, apart from the anti-environmentalism of the Reagan administration and Anne Gorsuch. I knew somebody at the University of Iowa who did discrete choice modeling. He invited me to visit the university for a few days. I had never been to Iowa City, but my wife had been there. She worked for the administrator of the nationwide physician's assistant training program in what was then the Department of Health, Education and Welfare. The University of Iowa's medical school had a grant to train physicians' assistants. My wife had been to Iowa City to review the medical school's program. So she knew Iowa City and was happy to move there. She eventually became the mayor of Iowa City.

*Did you go to the economics department?*

I was three quarters in geography and one quarter in economics.

*I guess transportation is closer to geography?*

Geography at the time wanted to become more quantitative. They were interested in discrete choice because of transportation and spatial choice. So I went to Iowa in 1982 with a 3/4-1/4 appointment in geography and economics, but eventually transferred to a full time appointment in the economics department.

*So that was after more than 10 years of a non-university environment?*

Yes. I was at the EPA for 11 years and at RAC for approximately four years before that. My career path has been unconventional and probably not one that today's younger people should try to emulate.

*And the location, in the DC area versus Iowa City?*

Moving to Iowa City was not a big shock. Iowa City is not rural Iowa. It is a small city, but it has a big university, good restaurants, and a rich cultural life. Nobody with life's normal obligations can attend all the plays, concerts, and other events that take place there. Iowa City is known internationally because of the university's creative writing and international writing programs. Tennessee Williams was a graduate of the University of Iowa. One of the university's theatres has a photo of him as a supernumerary in a play while he was a student. Nobody who knew him as a student expected him to amount to much. I guess forecasts of drama careers are no more accurate than forecasts of stock prices.

*Right, Iowa City is quite pleasant.*

Gary Fethke, who has a Ph.D. in economics from the University of Iowa, was in the management science department when I moved there, and we became friends. The economics department at the U of I is in the business school. The dean was George Daley, who was also an economist. Gary became the dean when Daley left to become the dean of the business school at NYU. So the economics department had a lot of high-level support despite being, arguably, an anomaly in the business school. Gene Savin came in 1986.

*Yes. So around that time, you became friends?*

Yes. Gene Savin, Bob Forsythe, a student named Martin Sefton, and I did some experiments with simple bargaining games and wrote a paper about their results. It is my most heavily cited paper, though I am not an experimental economist. Gene and I became good friends and wrote several other papers together in the following years. Gene was also enormously helpful to me as a mentor in my new career as an econometrician after all the twists and turns that had come earlier in my professional life. I also became friends and wrote several papers with George Neumann, who was the chairman of the economics department for several years. At some point, Gene and George suggested that I should move from three quarters in geography, one quarter in economics to three quarters in economics, one quarter in geography. Gary Fethke, who was the associate dean at that time, was also supportive. I changed from 3/4-1/4 to 1/4-3/4, but then Gene and the others said just move 100% into economics. So I moved full time to the economics department and remained there until I moved to Northwestern. The economics department was in the old business school building, Phillips Hall, when I became full time in the department. The building that now houses the department had not yet been built. Gary Fethke was deeply involved in the building's design and deserves a lot of credit for how nice the building is and for the artwork in it. More recently, he donated a modern sculpture by a living artist to the business school. It is displayed in front of the building.

*Speaking of this collaboration with your colleagues, this was not transportation. It was experimental economics.*

The paper with Forsythe, Sefton, and Savin is about experimental economics, but it had statistical content. We wanted to find out whether the results of certain kinds of experiments are replicable and consistent with the relevant game theoretic models. The results of experiments are random, so you need statistical methods for deciding whether they are replicable and consistent with game theory. It turned out that the experiments we did were replicable but not consistent with theory models. Most participants in the experiments did not try to maximize their own returns. Instead, they tended to divide the money involved roughly equally with their opponents, whom they did not know and could not see. The title of our paper is "Fairness in Simple Bargaining Games" [7].

*So that's very interesting. Kind of a junction of your research? Like, because at the time you're not doing that much theoretical econometrics.*

At that time, I was starting to work on what turned out to be the smoothed maximum score estimator. As you know, the objective function of Chuck Manski's maximum score estimator is a sum of indicator functions, which makes it hard to compute. It also causes maximum score estimates to have a slow rate of convergence and a very complicated limiting distribution. I had the idea of smoothing the indicator functions so they become differentiable. Under certain smoothness conditions that are stronger than Manski's, the resulting estimator has a faster (though not root-n) rate of convergence than the unsmoothed estimator and is asymptotically normally distributed. Also, the bootstrap provides asymptotic refinements for the smoothed estimator.

*I see. So you were doing first transportation then moving towards theoretical econometrics, through discrete choice?*

Yes. The smoothed maximum score paper was published in *Econometrica* [32]. I later learned that Chuck Manski had been one of the referees of the paper. Of course, he was an obvious choice.

*Did you know Chuck Manski?*

I knew Chuck before he got his PhD and have now known him for some 50 years. He was still a graduate student at MIT when we first met. EPA had a contract with one of the transportation faculty at MIT.  My EPA colleagues and I went to MIT from time to time to discuss the contract with the people who were doing the work. Chuck was in the economics department, but some of the most exciting research in discrete choice modeling was in transportation. Chuck came to some of the meetings about the contract and gave a seminar. That is how we met. Later, we wrote several papers together about partial identification. Also, Chuck is the person most responsible for my moving to Northwestern. I was an unusual case. I was 60 years old, did not have a Ph.D. in economics, and was in a much lower-ranked department.  Of course, I don't know what discussions went on at Northwestern before I moved there, but I am sure that Chuck made a strong argument for hiring me. Chuck also introduced me to David Pollard's book about empirical processes. I didn't know how to solve a problem I was working on. Chuck told me there was a book about empirical processes (which I had never heard of) that I should read.  That is how I learned about empirical process theory.

*So you met him a long time ago, but then you met Rob Porter?*

I met Rob Porter at the 1990 World Congress of the Econometric Society in Barcelona. He was the *Econometrica* co-editor who handled the smoothed maximum score paper. This was 11 years before I went to Northwestern. I also knew Joel Mokyr because we had been members of the NSF economics panel together. I had met Dale Mortensen when we were both visiting Aarhus University in Denmark. Dale had also attended an econometrics conference at the University of Iowa. Rosa Matzkin was also at Northwestern when I interviewed. I had met her when she was at Yale, so I had known her for a long time.

*So was Econometrica as important in that time as it is now?*

It definitely was. The smoothed maximum score paper was the lead article in its issue of *Econometrica*. I was happy that the paper had been published and didn't think being the lead article mattered much. It was like a random event. Something had to be the lead article, and I assumed that all the articles had equal probabilities of being the lead. However, it turned out to be a very big deal among my economics colleagues, which was a surprise.

*Was Chuck a kind of tech advisor at the time for you?*

Yes. He and Gene. Lars Muus was visiting Iowa and gave a series of lectures on something I had never heard of that he called the bootstrap. He said the bootstrap could do lots of wonderful things, and that set me to wondering whether the bootstrap could be applied to the smoothed maximum score estimator. While I was wondering, Wolfgang Härdle invited me to a conference at the Université Catholique de Louvain in Louvain-la-Neuve, Belgium. Chuck was there. So was Peter Hall, whom I had met and written a paper with when I visited the Australian National University in 1986. One day

Chuck and I took a train to Brugge (Bruges in French, but it is in the Flemish part of Belgium), which is an interesting medieval city. On the train back to Louvain-la-Neuve, Chuck told me about a problem he was working on. Suppose you want to know a parameter of the distribution of a random variable. You have a random sample that is sometimes drawn from the distribution of interest and is sometimes drawn from an unknown distribution. You don't know which distribution an observation was drawn from. We started talking about how to deal with this problem while we were on the train. We treated it as a partial identification problem and eventually wrote a paper about it. We submitted the paper to the *Annals of Statistics*. They desk rejected it on the grounds that the proofs were too simple for the *Annals*. Then we submitted the paper to *Econometrica*. Peter Robinson was the econometrics co-editor at the time. He liked the paper and accepted it for publication despite negative reviews by the referees. Unlike many journal editors, Peter didn't operate by a majority vote of the referees. He read the paper carefully and decided for himself whether he thought it was a good paper [39].

*I see. That's another good success story!*

**Experience as Co-Editors: Econometric Theory and Econometrica**

Tell us about how you met Peter Philips and became a co-editor of *Econometric Theory*.

In the fall of 1986, I had a semester of leave from the University of Iowa. I decided to go to the Australian National University to be a visitor at the Research School of Social Sciences. Gene Savin was visiting the University of Canterbury in Christchurch, New Zealand, at the time. I went to Christchurch on my way to Australia and visited Gene in his office at Canterbury. Peter Phillips, who is from New Zealand, stopped by while I was in Gene's office. I had never met Peter, but Gene introduced us and we talked for a while. In 1988, Peter invited me to become an associate editor of *Econometric Theory*, which he had started recently. In 1992, he asked me to become a co-editor. I was a co-editor of *Econometric Theory* until 2000, when I became a co-editor of *Econometrica*. I stopped being a co-editor of *ET* then for two reasons. One was because it was going to be too much work to be a co-editor of both journals at the same time. The other is that one person should not have control over which papers do or do not get published in two journals.

*Right, and in terms of editorial experience at ET, was it slower compared to the current standard? Or was it the same as now?*

It is hard to say. Electronic management of papers and the review process hadn't happened yet. Communication was on paper through the mail. Authors submitted several printed copies of their papers, and I sent printed copies to the referees. As a co-editor, I managed the review process and made recommendations to Peter about whether papers should be accepted or rejected. Peter made the final acceptance/rejection decisions. Peter never contradicted one of my recommendations, and I never received any pressure from him to keep the acceptance rate below a certain level. At *Econometrica*, some of the other co-editors felt that the acceptance rate should not be higher than 10%. Why 10%? I don't know. My acceptance rate at the time was somewhere in the teens. Some of the others said that was too high, but they did not say I was accepting bad papers. They just wanted my acceptance rate to be lower. Since then, the volume of submissions to *Econometrica* has increased and now the acceptance rate is something like 5%. It is harder than it used to be to get a paper accepted at *Econometrica* and other top journals. Many papers are desk rejected because of the volume of submissions.  A paper that goes through the normal review process may have many referees, and there can be several rounds of refereeing before a paper is accepted. I don't know whether this has improved the quality of the papers, but they have become longer. I think this is partly because of the need to satisfy many referees. The acceptance rate at *Econometrica* was below that of several top statistics journals when I investigated this several years ago. Glenn Ellison was the editor of *Econometrica* when I became a co-editor. He found the records of *Econometrica* from the 1940s. He told us that, conditional on acceptance, the mean time from submission to acceptance was 9 months and the mean number of rounds of review was less than one. So maybe econometrics is going in the wrong direction in that regard.

**Meeting Peter Hall and Collaborations with him**

*Let's go back and continue your story about how you met Peter Hall.*

After New Zealand, I went to the Australian National University (ANU) in Canberra to visit the Research School of Social Sciences. ANU had regular departments but it also had research schools where the faculty had no teaching responsibilities. I was a visitor and wouldn't have taught courses even in a regular department. At the time I was working on a paper about semiparametric estimation of censored linear regression models. There were a conditional quantile version and a conditional mean version. Both required nonparametric estimation of derivatives of the unknown cumulative distribution function of the model's random component. Estimation of the derivatives requires selecting a bandwidth. Bandwidth selection methods for kernel density estimation are not asymptotically optimal in this setting. I did not know how to select the relevant bandwidths. Peter Hall was at ANU then.  He later moved to the University of Melbourne.  I sent him an e-mail telling him who I was and asked if he would like to meet for lunch. He said yes, and we met at a restaurant on the ANU campus. I described the bandwidth problem to him.  He said it was a very interesting problem and that he would work on it "right now."  He had the ability to concentrate intensely on any problem he worked on, and he soon found a way to select the bandwidths for both models. This resulted in a co-authored paper that was published in *ET* [14].

I went back to ANU the following year.  During that visit, Peter and I discussed the problem of choosing the block length in the block bootstrap for time series. That led to a paper with Peter and Bing Yi Jing on block length in the block bootstrap. The paper was published in *Biometrika* [15]. It was my first but not my last paper with a co-author, Bing Yi in this case, whom I had never met, though I did meet him later during a visit to the Hong Kong University of Science and Technology. After the paper with Peter and Bing-Yi, I started working on using the block bootstrap to obtain asymptotic refinements in GMM estimation with time series data. That was a difficult problem. The theory of bootstrap asymptotic refinements involves Edgeworth expansions, and Edgeworth expansions for dependent data are more complicated than expansions for iid data. Eventually, Peter and I wrote a paper about asymptotic refinements for GMM with time series. It was published in *Econometrica* [13]. Later, Don Andrews refined our work by, among other things, relaxing some of our assumptions.

The next time Peter Hall and I had an opportunity to work together was when Peter Robinson invited me to give a talk at a statistics conference in Vilnius, Lithuania. The hotel in which my wife and I stayed had only a few rooms, and it turned out that Peter was staying in a room across the hall from us. Nonparametric instrumental variables estimation was becoming interesting in econometrics at the time. Peter thought it was interesting for statistics, too, and we agreed to work on it together.  Of course, it wasn't possible to make much progress during the conference, but we defined the problem we wanted to work on. Soon after I returned to the US, I got an email from Peter with several handwritten pages of work he had done on the nonparametric IV problem that we had defined.  He said he had done the work while flying over Afghanistan on his way back to Australia from Lithuania.

*So he was fairly good at working on new problems anywhere?*

Yes, including on the tiny table that you have on an airplane. We decided that the paper we were writing should include a motivation for nonparametric IV. I suggested using endogeneity of education in a wage equation. Peter didn't understand what I was talking about but said it was fine with him if I wrote that part of the paper. Peter was very interested in natural science and engineering, and he did applied work in those fields as a consultant for CSIRO (Commonwealth Scientific and Industrial Research Organization), but economics was a kind of religion for him, as it is for many statisticians and natural scientists.  We submitted our paper to the *Annals of Statistics*, which published it in 2005 [12]. The refereeing process at the *Annals* is different from the process at top economics journals. There are fewer referees, their reports are shorter and more to the point, and there are fewer rounds of review. The entire process takes months, not years.

After finishing the nonparametric IV paper, I started learning about functional regression, which has a long history in statistics and is relevant to certain problems in economics. I realized that the mathematical problem posed by functional linear regression is closely related to that of nonparametric IV, and the methods of nonparametric IV can be applied to functional linear regression. Both involve solving a Fredholm integral equation of the first kind, and both present ill-posed inverse problems. One way to estimate a functional linear regression model is to represent the unknown slope function and unknown covariance operator of the explanatory process (or variable) as series expansions with a user-chosen basis.  Another way is to use the eigenfunctions of the estimated covariance operator as the basis.  The latter method is widely used in statistics. Peter and I wrote a paper that gives conditions for consistency and rates of convergence for both methods. The rates of the two methods are the same, but the conditions needed to achieve them and for consistency are different. Our paper was published in the *Annals of Statistics* in 2007 [11].

*In econometrics at that time Jean-Pierre Florens and Xiaohong Chen got interested in nonparametric IV and they also contributed.*

Yes. Jean-Pierre had written a paper with Serge Darolles and Eric Renault. It took a long time to be published. In the end, they took on another co-author before the paper was published.

*Yes, Yanqin Fan joined them.*

Yes, it was Yanqin Fan. Peter's and my paper on nonparametric IV was written after but published before the paper by Jean-Pierre and his co-authors. Both papers use kernel methods and Tikhonov regularization. Nonparametric IV estimation is analogous to least squares estimation of a linear model in which the smallest eigenvalue of the $X'X$ matrix is arbitrarily close to zero. Tikhonov regularization is analogous to ridge regression and, roughly speaking, a way to prevent the near singularity of $X'X$ from causing the variance of the least squares estimate to explode. Another approach is to use series representations of the relevant functions and regularize through series truncation. This is analogous to replacing the $X'X$ matrix with another matrix whose smallest eigenvalue is further from zero. At roughly the same time that Peter's and my paper was published, Richard Blundell, Xiaohong Chen, and Dennis Kristensen published a nonparametric IV estimation paper that used series estimation and regularization through series truncation. Series methods are applicable to certain kinds of nonparametric IV estimation problems that cannot be treated with Tikhonov regularization, so the paper by Blundell, Chen, and Christensen was an important contribution that advanced the field.

Peter moved from ANU to the University of Melbourne in 2006. I visited him there in 2011, and we began work on another paper. Nonparametric estimation of a conditional mean function by a kernel method yields an asymptotically normal but asymptotically biased estimate. The asymptotic normal distribution is not centered at zero; it is centered at the expected value of the estimate, which is not the true conditional mean function. Series estimation also yields asymptotically biased estimates. Some people have suggested ignoring the asymptotic bias when obtaining confidence intervals. This always results in under-coverage of a symmetrical confidence interval as a consequence of the algebra of the normal distribution (or shape of the normal probability density function). Of course, this was not a new problem. There were two well-known methods for dealing with it.

One method was under-smoothing. The other was explicit bias reduction. Both methods require an auxiliary bandwidth, and there was no empirical method for choosing this bandwidth. Moreover, even with the auxiliary bandwidth that minimizes the error in the coverage probability, the coverage probability is much too low in samples of reasonable size. The idea, which was Peter's, that we started working on was: Suppose you ignore the bias but widen the confidence interval by making its critical values large enough to overcome the under-coverage problem. Of course, if the interval is wide enough, the coverage probability is one, but you don't want that. It was also known that bootstrap sampling from the empirical distribution of the data is inconsistent because the bootstrap estimates the distribution of the expected value of the kernel estimator, not the distribution of the properly centered estimator.

These problems led us to ask whether there is a bootstrap method for correctly choosing the critical values for the widened confidence interval. The answer is yes. Nadaraya-Watson and local linear estimators can be well approximated as the sum of a mean-zero stationary Gaussian process and a non-stochastic term. The same is true for the bootstrap estimate. If you could do a Monte Carlo simulation in which you sampled the true process, you could have many simulated observations at each X value and use these to form a confidence interval for the conditional mean function. But you cannot do this simulation in an application because you do not know the true Gaussian process. However, the Gaussian process is stationary and oscillates very rapidly. Therefore, sampling along the process instead of at a fixed point is like sampling at a point. In terms of statistical physics, it is like taking a time average instead of an ensemble average. You can use the bootstrap to do this and, thereby, choose the critical values for the widened confidence interval to minimize the probability of under-coverage. This works much better than undersmoothing or explicit bias reduction. Peter and I wrote a paper about this method that was published in the *Annals of Statistics* [10]. It was our last paper. Peter became seriously ill and died shortly after the paper was published. Several years later, Sebastian Calonico, Matias Cattaneo, and Max Farrell found a way to do explicit bias reduction that makes somewhat stronger assumptions than Peter's and my method but does not require an auxiliary bandwidth and works well in finite samples.

Peter and I discussed extending our method to nonparametric estimation of a conditional quantile function. The cusp in the objective function for quantile estimation creates complications that do not arise with conditional mean functions. We did not get a chance to work on the extension before Peter became ill. My student, Anand Krishnamurthy and I carried out the extension and published it in a special issue of *Statistica Sinica* in honor of Peter [34].

*You had more than two decades of a very productive joint collaboration with Peter.*

Yes, for about 25 years. It started in the late 1980s and went on until about 2013.

**Collaborations with George Neumann**

*Going back to the late 80s, around that time another colleague of yours was at Iowa, George Neumann.*

At that time, I was working on semiparametric estimation. Semiparametric estimation was relatively new in econometrics, and there was a lot of interest. Many people were working on it. There was an *Econometric Reviews* session at the Summer Meeting of Econometric Society. The editor of *Econometric Reviews*, Esfandiar Maasoumi, invited me to give a talk at the meeting, which was going to be held at UC Berkeley. I had an idea about how to estimate a duration model that was not a proportional hazard model but contained it as a special case.  However, I didn't have a good way to motivate my model or explain why anyone would want to use it.  George knew duration modeling problems that arose in the analysis of income maintenance experiments that had been done in the late 60s and early 70s, and he had relevant data. We decided to work together on an application that was based on the income maintenance experiments. We presented the paper at the Summer Meeting, and it was published in *Econometric Reviews* [43].

The Cox proportional hazard model is a well-known and widely used duration model. George and I started discussing the question of how you can test the hypothesis that the Cox proportional hazards model is correctly specified. We wrote a paper about how to test that hypothesis against a nonparametric alternative.  That paper was published in the *Journal of the American Statistical Association* [40].

*I recall that after you had done functional data work, you wrote another paper with George Neuman along with Federico Bugni and Peter Hall.*

Peter and I had discussed the following problem.  Suppose you have data consisting of realizations of a stochastic process. The realizations are functional data. Suppose you also have a model that generates a stochastic process.  How can you test the hypothesis that the model generates the observed process?  Federico, who was a student of mine at the time, Peter, and I had a method for doing this test, and we wanted an empirical application.  George, Dale Mortensen, and Ken Burdett had developed a model of the wage generation process in labor economics.  Moreover, George had wage data that could be used to test the hypothesis that the data were generated by the model. Our test rejected the model. I was worried about this, because George is a co-author of the paper that describes the model we tested. I thought he might be upset with the result, but he wasn't.  He said, "If that's the way it turns out, then that's the way it is. Science has to move ahead based on facts, not what is convenient for people."  George was very accepting of this.

*George had a physicist's kind of attitude about theory and empirics, then?*

Yes.  Our paper was published in the *Econometrics Journal* [5].

*Did you write other papers about functional data?*

Yes.  Federico Bugni and I wrote a paper that was motivated by experiments in pricing electricity and gas in which households were assigned randomly to treatment groups with different pricing and billing schedules and a control group. Electricity and gas are consumed in continuous time and can be measured frequently (e.g., every 30 minutes), so observations of consumption over time are functional data. We wanted to test the null hypothesis that the stochastic processes generating the functional data were the same for the treatment and control groups. We developed a nonparametric permutation testing procedure for this purpose and applied it to data from an experiment on gas pricing in Ireland. The paper was published in the *Journal of Applied Econometrics* [6].

**Collaborations with Vladimir Spokoiny**

*Speaking of specification testing, one thing that comes to my mind is your work with Spokoiny about testing a parametric mean-regression model against a nonparametric alternative.*

Yes, that happened in the early 2000s. Wolfgang Härdle arranged for me to get an Alexander von Humboldt Award for Senior U.S. Scientists that enabled me to visit Wolfgang each year at the Humboldt University in Berlin. I met Vladimir (Volodia) Spokoiny during one of these visits. He is a Russian mathematical statistician whose main appointment is at the Weierstrass Institute for Applied Analysis and Stochastics, which is within walking distance of the university. Wolfgang and Enno Mammen had developed a bootstrap method for testing a parametric model of a conditional mean function against a nonparametric alternative. Their paper was one of the first to use the wild bootstrap. Their method is a pointwise test in the sense that there is a single true function with a specified smoothness. Volodia and I wanted to develop a test that is consistent uniformly over a smoothness class. Functions that are in the smoothness class but near its boundary can be very wiggly. Consequently, the asymptotic distribution of a pointwise test can be a poor approximation to the finite-sample distribution unless the sample is very large. This is analogous to testing a hypothesis about a population mean that is known to be greater than or equal to zero. If the population mean exceeds zero, the pointwise asymptotic distribution of the Student t statistic is normal, but the finite-sample distribution can be far from normal if the true mean is close to zero. The first question Volodia and I wanted to answer was, "Is there a test that adapts to the unknown smoothness of the nonparametric alternative and is consistent with the fastest possible rate of testing uniformly over the alternative's smoothness class?" The rate of testing is the rate at which the distance between the parametric model and nonparametric alternative converges to zero. Uniform consistency means, roughly speaking, consistency against the worst-case alternative in the smoothness class. Worst case alternatives are very wiggly, which can slow the rate of testing. Moreover, worst case alternatives cannot be represented as sequences of local alternatives. This led to our second question: Can a uniformly consistent test achieve a rate of testing against a sequence of local alternatives that is nearly $n^{-1/2}$? We developed an adaptive test that achieves the fastest possible uniform rate of testing and has a rate of $n^{-1/2}\sqrt{\log\log n}$ against a sequence of local alternatives. Our paper about this test was published in *Econometrica* [45]. We wrote a second paper that was about testing a linear quantile regression model against a nonparametric alternative. That paper was published in the *Journal of the American Statistical Association* [44].

**Visits to Germany and Collaborations with Wolfgang Härdle**

*We have not yet covered all your research to the mid-90s, so maybe we go back a bit. Let's we talk about your frequent visits to Germany in 1990s.*

I met Wolfgang Härdle at a conference at the Catholic University of Louvain (UCL in French) in Louvain-la-Neuve (LLN), Belgium. Wolfgang was on the faculty there. His home was in Bonn but he rented a house in LLN, which he shared with several graduate students. I returned to LLN the following year and met Alexandre (Sasha) Tsybakov. He is a Russian mathematical statistician like Volodia. He is now at CREST-ENSAE in Paris and is also a professor at the University of Paris. Once he took me into a closet in his office and pointed at a box of loose papers with mathematics written on them. They were Maurice Fréchet's handwritten notes. I also met Irene Gijbels and Léopold Simar, among many others. I returned to LLN several times in the years after the conference.  Wolfgang and I got to know each other better, which led to our writing several papers together. Sometime after that, I was invited to spend several weeks each year at Tilburg University in the Netherlands. The econometrics department there had an apartment in Tilburg that I used. Wolfgang got a visiting position at Tilburg, and we shared the apartment. He also had a fellowship at Humboldt University in Berlin in addition to his position at UCL, so he was busy. He spent weekends in Bonn with his family.  I went with him to his home in Bonn once and we started talking about the problem of testing a parametric model against a semi-parametric alternative. This was before Volodia and I began working on our test against a nonparametric alternative. Wolfgang and I continued our discussions after we returned to Tilburg. These led to a paper about a test of a parametric model against a semiparametric alternative. That paper was published in *Econometric Theory* [17].

Sometime after that, Wolfgang took a permanent position as the professor of statistics at Humboldt University, and I began visiting Berlin instead of Tilburg. Wolfgang shared a suite of offices with Helmut Lütkepohl, who was the professor of econometrics. Wolfgang once took my wife and me to see the house in Berlin he hoped to buy. It was on the far eastern border of the city near the Spree River and came with a small sailboat. The owner was the widow of a professor of agriculture at one of the universities in Berlin. Buying the house took a long time because there had to be an investigation of whether it had been confiscated from a Jewish family by the Nazis. It hadn't been.  Wolfgang was able to buy the house and fixed it up on his own.

*Did your visits to Berlin take place in the summer?*

Yes.  In principle, a recipient of an Alexander von Humboldt Award was supposed to go to Germany for a year. However, Wolfgang and I decided that I would go for a month or so each year for several years. This enabled me to meet many people I would not have met otherwise, including Spokoiny and Enno Mammen with whom I developed research collaborations. Gerard Debreu was a visitor one year. He was a courtly man and traveled to Zagreb frequently to visit a friend, despite the war that was going on in Yugoslavia at the time.

*So eventually you became a multiyear visitor?*

Yes. At that time there were two well-known methods for estimating the coefficient vector in a semiparametric single index model of a conditional mean function. One was Hidehiko Ichimura's semiparametric nonlinear least squares estimator. The other was average derivative estimation, possibly with density weighting. Ichimura's estimator is hard to compute because it requires solving a non-convex optimization problem that can have many local optima. Average derivatives are much easier to compute but are defined only for continuously distributed covariates. At least one continuous covariate is necessary for identification, but many applications have discrete covariates as well as continuous ones. Wolfgang and I wondered whether it was possible to develop a method that accommodated discrete covariates but had the computational advantages of average derivative estimators. It is possible. The coefficients of continuous covariates can be estimated by average derivatives. The estimated conditional mean function is a smooth function of the continuous covariates whose intercept depends on the discrete covariates. Changes in the discrete covariates change the intercept and, thereby, the location of the function but leave its shape unchanged. The coefficients of the discrete covariates can be estimated consistently from the changes in the location. Wolfgang and I wrote a paper describing how this is done. We also wrote code to implement the method. The paper was published in the *Journal of the American Statistical Association* [16].

*So how was working with him? Was he kind of a more detailed person or a big picture person?*

He was a big picture person. Humboldt University had an old-fashioned German faculty structure, and professors like Wolfgang had a great deal of power. He supervised a large group of lower ranked faculty, graduate students, and post-docs who were working on their habilitations (a kind of second dissertation that existed in the German system). He also had visitors like Mammen and Debreu, among many others.

*So your collaboration with Wolfgang was successful.*

Yes. At some point I started doing work about the bootstrap. Wolfgang had instituted an annual lecture that he called the Herman Otto Hirschfeld Lecture. Hirschfield was a German statistician who moved to England in the late 1930s and changed his name to H.O. Hartley. He moved to the United States in 1953. I was the first Hirschfeld lecturer. The lecture was about the bootstrap. I later gave short courses on various bootstrap topics, including the block bootstrap for time series, in London, Aarhus, and Heidelberg. Through my visits to Berlin and my work on the bootstrap, I met Jens-Peter Kreiss. He is a statistician at the Technical University of Braunschweig who has done a lot of research on time series and frequently visited Wolfgang's institute in Berlin. The three of us decided to write a review paper about bootstrap models for time series. It was published in the *International Statistical Review* [18].

Wolfgang and a group of French statisticians organized an annual conference called the Paris-Berlin Seminar. It met for a week each year and alternated between locations in France and Germany (not necessarily Paris and Berlin). In addition to the usual talks, two people were invited to give short courses that met for one hour each day. Wolfgang invited me to give one of the short courses. The seminar that year was at a disused geophysical observatory in a rural region (the Sancerre wine region) of the upper Loire valley. My course was about semiparametric methods in economics. The audience was French, German, and Russian mathematical statisticians. I was the only American there and, more importantly, the only non-mathematician. I was afraid that these people would find my lectures trivial and uninteresting, but the Germans and Russians turned out to be very interested and asked lots of questions. It was obvious to them that I am not a mathematician, but they could see interesting mathematical problems in the topics I discussed. When the week was over, I went to the railroad station and stood next to Lucien Birgé while waiting for the train to Paris. He turned out to be an expert on Parisian restaurants and made some good recommendations.

Later, I turned the Paris-Berlin lectures into a book that was published in the Springer lecture notes series [27]. Much later than that, I expanded the Paris-Berlin book into another book that treated a wider range of topics in semi- and nonparametric estimation. It was published in the Springer series in statistics [23].

**Research on Transformation Models**

*Around that time, you had papers about non-parametric and semi-parametric transformation models. Can we talk a bit about that?*

That grew out of work on duration (or survival) models that I did in the 1990s. It began with a paper about nonparametric estimation of the transformation function in a linear transformation model whose random error term has an unknown distribution. The transformation function is a functional of the distribution of the dependent variable conditional on the covariates. The functional can be estimated nonparametrically using standard methods. I wrote a paper that described the details of the estimation procedure and the asymptotic properties of the estimated transformation function [29].

I then turned to the Cox proportional hazard model with unobserved heterogeneity. The proportional hazard model was widely used in labor economics, for example to model the length of a spell of unemployment. Jim Heckman and Burton Singer had written a paper showing that if you have a proportional hazard model with unobserved heterogeneity that is ignored, the sign of the slope of the estimated baseline hazard function can be incorrect. This was important in labor economics at the time because one of the open questions was whether long-term unemployment was stigmatizing or, alternatively, led to a more vigorous search for employment. Stigmatization could cause the baseline hazard function to have a negative slope, whereas a more vigorous search could cause the slope to be positive. The proportional hazard model with unobserved heterogeneity can be written as a transformation model. The transformation function is unknown, and the random error term is the sum of a random variable with an extreme value distribution and a random variable with an unknown distribution.  My first paper showed how to estimate the transformation function. The estimated transformation function is related but not identical to the baseline hazard function of the proportional hazard model, and the random error term is related but not identical to the convolution of the extreme value and unobserved heterogeneity distributions.  I wrote another paper that explained how to estimate the baseline hazard function and distribution of the unobserved heterogeneity. The paper also presented the asymptotic properties of the estimators [26].  Both papers were published in *Econometrica*.

**Smoothed Maximum Score Estimator**

*Your smoothed maximum score was your first big breakthrough in your career. After that, you had multiple papers in Econometrica, and quite soon you would become a co-editor of Econometrica.*

There were two smoothed maximum score papers in the early 90s.  The first one was theoretical and was finished before the conference in LLN.  It is the *Econometrica* paper [32].  In the second paper, I applied the smoothed maximum score estimator to a real data set and compared its results with the results obtained with several other semiparametric and parametric methods.   I presented that paper at the conference in LLN and published it in the *Journal of Econometrics* [24].

Chuck Manski introduced the maximum score estimator in 1975 and described its properties in several subsequent papers. The maximum score estimator for a binary response model is the binary analog of a linear quantile regression.  In a 1990 article in the *Annals of Statistics*, Jeankyung Kim and David Pollard showed that the rate of convergence of the maximum score estimator for a binary response model is $n^{-1/3}$. They also showed that the asymptotic distribution of the suitably centered and scaled estimator is the maximum of a complicated and highly intractable multivariate Gaussian process. Chris Cavanagh obtained a similar result in an unpublished paper. The complexity of the analytic asymptotic distribution precludes its use for inference in applications. Jason Abrevaya and Jian Huang showed that the bootstrap does not estimate the asymptotic distribution consistently.

The reason for the slow rate of convergence, complicated asymptotic distribution, and bootstrap inconsistency is that in the maximum score objective function, the parameter to be estimated is in the argument of an indicator function. The discontinuity of the indicator function is the source of the trouble.  If the indicator function is replaced by a function that increases from 0 to 1 smoothly, then the usual Taylor series methods of asymptotic theory can be used to derive the asymptotic properties of the resulting estimator, which is the smoothed maximum score estimator. If the smooth function that replaces the indicator function is fixed, then the smoothed estimator is not consistent for the parameter of the maximum score model. Consistency can be achieved by replacing the indicator function with a sequence of smooth functions that converges to an indicator function as the sample size approaches infinity. This complicates the derivation of the estimator's asymptotic distribution, but the result is an estimator that converges more rapidly than the original maximum score estimator and is asymptotically normally distributed. Of course, the smoothed estimator requires stronger assumptions than the unmodified maximum score estimator. These assumptions are mainly about the smoothness of the distributions that generate the data. Under these assumptions, the smoothed estimator depends on observations within a neighborhood of the parameter value at which the indicator function jumps from 0 to 1, whereas the unmodified estimator uses only observations that are arbitrarily close to the knife edge. If the distributions that generate the data are sufficiently smooth, the smoothed estimator achieve nearly an $n^{-1/2}$ rate of convergence. These results are presented in the *Econometrica* paper [32].

An extension of this work showed that the bootstrap provides asymptotic refinements for test statistics and confidence intervals based on the smoothed maximum score estimator. The bootstrap is inconsistent in unsmoothed (standard) maximum score estimation because of the non-smoothness of the maximum score objective function. In smoothed maximum score estimation, the objective function is smooth by construction and has derivatives. This makes it possible to do Edgeworth-like asymptotic expansions of statistics based on the smoothed maximum score estimator and its bootstrap analog. These expansions show that the bootstrap provides asymptotic refinements, though not of the same order as in smooth parametric models. This result was published in the *Journal of Econometrics* [24].

Another extension was motivated by a paper by D. De Angelis, Peter Hall, and G.A. Young, which had been published in *JASA*. It was about applying the bootstrap to a linear a median regression model estimated by least absolute deviations (LAD). Statistics based on the LAD estimator do not have conventional asymptotic expansions and the bootstrap does not provide asymptotic refinements because the LAD objective function is not smooth enough. It has cusps. In addition, a Studentized test statistic involves nonparametric density estimation. De Angelis, Hall, and Young showed that the rate of convergence of the error of the bootstrap approximation to the distribution of a Studentized test statistic is very slow. It turns out, however, that smoothing the cusps in the LAD objective function has the same effect as smoothing the discontinuities in the maximum score objective function. Specifically, the bootstrap provides asymptotic refinements for test statistics and confidence intervals based on the LAD estimator. I wrote a paper about that. It was published in *Econometrica* [28].

**The First Visit to London**

*Since you mentioned Richard at UCL in London, maybe we should talk more about your first time in London.*

One year in the late 1990s, Gene Savin organized the summer meetings of the Econometric Society at the University of Iowa. The summer meetings attracted relatively few senior people, so Gene decided to invite some senior people to give plenary lectures. He wanted to give younger people an opportunity to be exposed to senior people and to give senior people an opportunity to find out what younger people were doing. Richard Blundell was one of the senior people Gene invited. I didn't know Richard at the time, but he made several visits to the University of Iowa in the late 1990s.  Gene introduced me to Richard and said Richard should invite me to London. So Richard invited me.

*Had Gene been in the UK for a long time?*

Gene had been at Trinity College Cambridge for a long time.

*I guess he knew many British academics?*

Yes. Gene got his Ph.D. at Berkeley and had been at UBC and Northwestern in addition to Cambridge, so he knew a lot of people.

*Then Richard followed Gene's advice and invited you, right?*

Yes.

*So that was the late 90s?*

Yes. I had started working on nonparametric instrumental variables estimation. Nonparametric IV estimates converge much more slowly than nonparametric estimates that do not involve endogenous explanatory variables.  Richard and I decided that it would be worthwhile to have a nonparametric test for exogeneity of the explanatory variables against the alternative of endogeneity. We developed such a test and submitted a paper describing it to the *Journal of the American Statistical Association*. It was desk rejected on the grounds that it was about instrumental variables and therefore not statistics. Chuck Manski was outraged by this and wrote a letter of complaint to *JASA*. Richard and I submitted the paper to the *Review of Economic Studies*, and they accepted it [1].

*Statistics nowadays probably would not reject the paper based on that IV problem.*

JASA has many associate editors. I guess our paper was sent to one who didn't like IV estimation. At roughly the same time, Peter Hall and I were writing our paper about nonparametric IV estimation. We sent it to the *Annals of Statistics*. They didn't have any complaints about instrumental variables and accepted the paper after we made some reasonable revisions [12].

**Moving from Iowa to Northwestern**

*In terms of your Iowa time, I don't know whether we missed anything but you moved after your long period at Iowa and decided to join Northwestern around 2001. Can we talk a little about that decision and its background?*

When Chuck Manski was at the University of Wisconsin, he asked me if I would be interested in an offer from Wisconsin, and I eventually got one. However, my wife had become mayor of Iowa City. Moving to Madison at that point was not an option, so I declined the offer. Soon after that, Chuck moved to Northwestern. After he had been there a few years, he asked me if I would be interested in an offer from Northwestern. My wife's term as mayor had ended, so we could move to Evanston if we wanted to. I knew a lot of people at Northwestern in addition to Chuck, including Dale Mortensen, Rosa Matzkin, and Rob Porter. My job talk was about the uniformly consistent test of a parametric model against a nonparametric alternative that Volodia Spokoiny and I had developed. The talk must have gone well, because I eventually got an offer from Northwestern and moved there.

*I was a student at Iowa when you moved to Northwestern - I was almost finished. You had long-term colleagues like Gene Savin and George Neumann. Were they supportive of your decision?*

Yes, and John Geweke, whose office was next to mine. Gene's immediate reaction when I told him about the Northwestern offer was, "You should accept it." There was no discussion at all. George and John said similar things. The economics department at Iowa is in the business college, and Gary Fethke was the dean. He tried to persuade me to stay at Iowa, but I don't think his heart was really in it. He understood the attraction of Northwestern.

*Before we move to your time at Northwestern, is there anything you would like to add to your account of your time at Iowa? If I think about it, you moved to Iowa in 1982 and only had a part time appointment initially for the department of economics. Then, you switched to full time around 1987. I guess there is an enormous difference between your starting point at Iowa and your departure point. I mean, in the beginning, one could say you might not be called an econometrician, but more a transportation researcher. Then, when you left in 2001, you had not only become an econometrician but a theoretically oriented econometrician, so a huge change! But as you said, on the first day of the interviews, it's all based on your previous work.*

On my work as a theoretical physicist, right.
*So this is a full circle.*

**Joining Northwestern in early 2000**

*Let me ask you about your memories about Northwestern when you joined the department.*

I went there in September of 2001. I was co-editor of *Econometrica* and had a quarter time appointment with the Transportation Center. Francesca Molinari was at the end of her third year as a graduate student when I arrived. Chuck Manski was the chairman of her committee, and I became a member. She and I had many discussions about the work she was doing on partial identification in the presence of missing or corrupted data. I also spent quite a lot of time each week on *Econometrica*. *Editorial Express* hadn't yet begun. Correspondence with authors, referees, and the managing editor took place by postal mail. There were physical files of referee assignments, schedules, and reports. Eventually, *Editorial Express* replaced postal mail and physical files. The economics department was in the old Kellogg Building at the time. Rosa Matzkin, Chuck Manski and Dale Mortensen were there. Elie Tamer had not yet come. Bruce Meyer, Joe Altonji, and Chris Taber were also there. There were also a lot of junior people. Northwestern's procedures for making third year reappointment, tenure, and promotion decisions were very different from Iowa's, so I had to learn how the system at Northwestern works.

*Did you have any junior econometricians at the time?*

We had an assistant professor, Luojia Hu, a former student of Bo Honoré's. She eventually left for the Chicago Fed and is still there. The other senior econometricians were Rosa and Chuck. Richard Spady was an adjunct professor. He is now at Johns Hopkins.

**Frequent Visits to University College London**

*So, you moved to Northwestern, and you were a frequent visitor to Europe, notably to London and University College London (UCL). Now we are in London, so perhaps this is a good time to talk about your time in London.*

I started going to London in the late 1990s. Richard Blundell and I started working on various problems, and you came a few years later.

*Right, I joined UCL in 2002.*

You and I worked on several papers as well. And at the time, I was getting increasingly interested in nonparametric IV. Richard was very interested in how to use innovative econometric methods and was quite receptive to these ideas. Our first paper was the one about testing for exogeneity in nonparametric IV. The null hypothesis is that the explanatory variable in the nonparametric function is exogenous. The alternative is that it is endogenous. An obvious Hausman-like way to test this hypothesis is to estimate the structural function under both hypotheses and use some statistical procedure to compare the estimates. If they are different by more than can be explained by random sampling error, the null hypothesis is rejected. But estimation under the alternative hypothesis has to be done by nonparametric IV, and nonparametric IV estimators have rates of convergence that are slower than the rates of convergence of nonparametric estimates under the null hypothesis. So the Hausman-like test produces two estimates that converge at different rates. The reason for this is the ill-posed inverse problem of nonparametric IV estimation. The question that Richard and I had to deal with was, "Is there a way to do the test that avoids ill-posedness?" We found a way to do this. The REStud paper that we talked about earlier describes the resulting test and its properties [1].

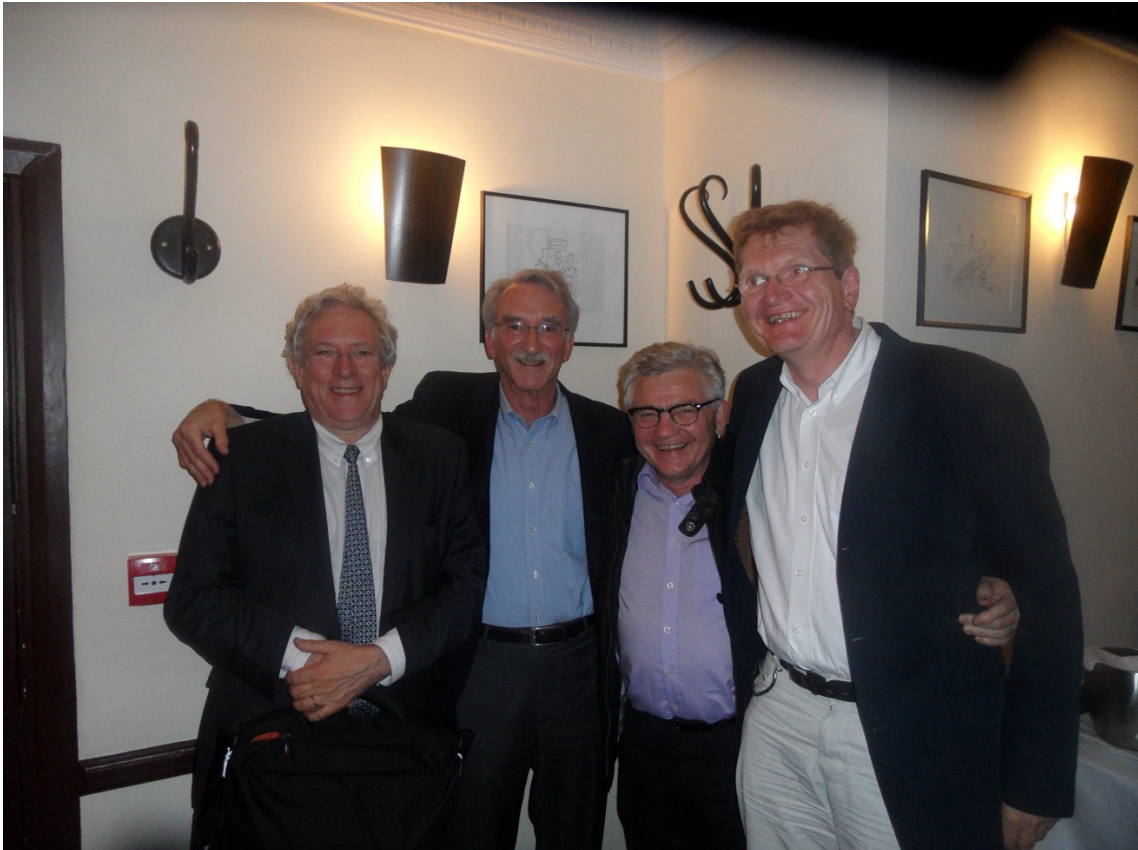

(Photo caption: Andrew Chesher, Joel Horowitz, Wolfgang Härdle, Enno Mammen (from left to right), Dinner at the Giaconda Dining Room, London, June 2011)

## **CEMMAP**

*You also participated in CEMMAP activities. Andrew Chesher said you gave the first CEMMAP masterclass.*

I was the first. I gave several others after that. For a while, I had given more master classes than anybody else, but Chuck Manski has now overtaken me by a considerable amount.

*Perhaps we can talk about the initial environment at CEMMAP.*

*CEMMAP had not yet been founded when I first started going to London. It was founded in 2000 with a grant from the Leverhulme Trust. I and many other people became CEMMAP International Fellows. Master classes helped to pay for my trips to London. I would teach one of the classes for two days as part of a longer visit.*

*The first master class took place in April 2002. The topic was "Bootstrap Methods in Econometrics", which I guess was based on your bootstrap handbook chapter. Was this chapter written before you joined Northwestern?*

I was still in Iowa. Jim Heckman and Ed Leamer were the editors of the volume. Jim invited me to write the chapter about the bootstrap. Instead of having the usual refereeing process, Jim and Ed had a conference at which people who had been invited to write presented their material. The number of people who had been invited exceeded the number of chapters Jim and Ed expected to publish. Each presentation lasted 30-45 minutes. Then there was a discussion of the presentation. There was also a follow-up refereeing procedure. I was asked to review a chapter that somebody else had written. Eventually I got comments and suggestions from Jim and Ed about my chapter. I revised the chapter as they suggested, and it was published in volume 5 of the *Handbook of Econometrics* [25]. Nearly 20 years later I was invited to write an article about the bootstrap for the *Annual Review of Economics* [20].

*Let's talk about your CEMMAP master class.*

*CEMMAP* was new at the time, and Andrew Chesher had the idea that *CEMMAP* should have master classes. It was a way to publicize and, perhaps, help finance *CEMMAP*, among other things. The master classes covered a variety of topics that were active research areas in econometrics. They still do. Andrew invited me to give the first class. We decided that it should be about the bootstrap. This was around the time that I had written several bootstrap papers with Peter Hall. I also had a paper about the information matrix test, which was not with Peter Hall, and I had been working on the handbook chapter about the bootstrap. A lot of people in statistics had done research on the bootstrap in the 20-some years since Brad Efron first described the idea in the late 1970s. In econometrics, however, there were few people who understood the bootstrap or were doing research on it. I was doing research on the bootstrap and its usefulness for applied econometrics. The master class lasted for two days and was well-attended. It covered topics that are in the handbook chapter, such as what the bootstrap is; what is the sense in which it works; and under what conditions it does or does not provide a consistent estimate of the asymptotic distribution of a statistic. Then we talked about asymptotic refinements, what asymptotic refinements are and conditions under which the bootstrap provides them. The class also included the use of the bootstrap with nonparametric methods such as kernel nonparametric density and regression estimation. It has somewhat different properties in nonparametric estimation because of asymptotic bias. We talked about the bootstrap for time series, and there were numerical examples illustrating finite-sample performance of the bootstrap. Subsampling was another topic. Subsampling overcomes all the known inconsistency problems of the bootstrap, provided that the object of interest has an asymptotic distribution, but subsampling is not a free lunch. The rate of convergence of its approximation errors is slower than those of the bootstrap when the bootstrap is consistent. Also, one must choose the subsample size to use subsampling in application. So, there is a sense in which subsampling is a last resort. Later, Peter Hall provided an example in which subsampling obtained a kind of asymptotic refinement, though not as good as the refinement that the bootstrap could achieve if the conditions in which the bootstrap could be used were satisfied.

Federico Bugni, who was my student at the time and is now a colleague, was interested in applying the bootstrap to moment inequality estimators. Victor Chernozhukov, Han Hong, and Elie Tamer had published a paper in which they showed that the conventional bootstrap applied to moment inequalities is inconsistent. They proposed using subsampling to overcome this problem. Federico became interested in the question of whether there was a way to make the bootstrap work with moment inequalities and, if there was such a way, was it more or less accurate than subsampling. Federico found a way to make the bootstrap work in moment inequality problems and was also able to show that the bootstrap achieved a faster rate of convergence to the asymptotic distribution than subsampling did. This work turned into his job-market paper, and a version of it was published in *Econometrica*.

*So, keeping off the path a bit, was Elie a faculty colleague?*

Elie came to Northwestern in 2004 and was a member of Federico's thesis committee.

*I remember the conference organized by Elie about partial identification when maybe Federico was a student. It was a huge conference, and all the important people in partial identification came and gave nice talks.*

Yes. We had several conferences, including one on partial identification and another on shape restrictions.

*That is a bit later. And your colleague, Ivan Canay, was on the job market at the same year as Federico. Northwestern hired him that year as an assistant professor.*

Ivan came from the University of Wisconsin. He was in the same year but arrived after Federico had left. They are both Argentinian (though they are now US citizens) and work together. Federico is now a member of our faculty.

**Nonparametric Instrumental Variable Estimation**

*Tracking back to CEMMAP, a few years later you gave this second masterclass about nonparametric IV problems.*

Yes. There was a lot of interest in nonparametric IV in econometrics, and it was becoming quite visible. A similar set of problems had been around in statistics and mathematics for roughly a century. The nonparametric IV estimation problem is a version of solving a Fredholm integral equation of the first kind. CT scanning solves a certain Fredholm integral equation of the first kind. Johann Radon discovered some of the mathematics for doing this (the Radon transform) in 1917. Certain problems in image denoising and other applications also involve solving Fredholm equations of the first kind. Analytic solutions to these problems are not available, and numerical methods can be very delicate. So problems related to nonparametric IV estimation had a long history in mathematics and statistics, but the problems addressed in those fields are not the one that arises in nonparametric IV estimation. The identifying relation in nonparametric IV includes an operator that must be estimated from the same data that are used to estimate the structural function. This is a source of significant complexity that was not present in the problems that mathematics and statistics literature addressed. The solution to the nonparametric IV problem in econometrics has properties similar to the solutions of the mathematics and statistics problems, but solving the econometric estimation problem is harder. This made nonparametric IV a popular topic in econometrics, but relatively few people were working on it because few econometricians were familiar with the necessary mathematical methods. So, I gave a *CEMMAP* masterclass on nonparametric IV. I also gave short courses on the bootstrap and nonparametric IV in Aarhus, Denmark and Heidelberg.

*I don't remember the exact timing, but you also gave a keynote speech at the European Meeting of the Econometric Society on nonparametric IV.*

Yes. Thierry Magnac, who was the program chair of the European Meeting in 2008, invited me to give the Fisher-Schultz lecture. The meeting that year was in Milan. The lecture was about nonparametric IV, but it was much shorter than a master class or other short courses.

*Thinking about this whole topic, as you said, it was very popular at the time and had big breakthroughs in terms of research. But there's not much follow-up nowadays. What's your prediction about this topic and potential impact on applied work?*

So far, it has had little impact on applied work. I think there are several reasons for this. One, which is not specific to nonparametric IV but applies to a lot of nonparametric methods, is that there is resistance to these methods among many applied researchers. This is partly because non- and semiparametric methods cannot treat the complicated structural models used in some applications fields. Another reason is that once you move away from standard nonparametric regression methods, the theory can get rather complicated. The theoretical work in these areas often proceeds more slowly than work in applications areas. Consequently, people doing current applied research often say that econometricians are working on a class of problems that applied people dealt with many years ago. Something else, which pertains specifically to nonparametric IV, is that most of the nonparametric IV literature is about continuous endogenous regressors and continuous instruments, whereas the endogenous variables and instruments are discrete in much (maybe most) of the applied IV literature. With discrete variables, the structural function is not necessarily nonparametrically point identified, and the theory developed in the nonparametric IV literature is not applicable. Even if the relevant variables are continuous and the conditions for point identification are satisfied, nonparametric series estimation of the structural function requires truncating the series. The resulting estimation problem looks like a finite dimensional parametric IV problem and can be solved using the standard methods for linear IV estimation. Inference is different and more complicated, however. Because the nonparametric IV series estimation problem looks like linear IV, it is easy to ignore the complications of nonparametric inference and proceed as if the estimation problem were truly parametric. Because of these and probably other reasons, nonparametric IV doesn't get used much in applications these days. I won't make predictions about how nonparametric IV might affect applied research in the future. A well-known aphorism is that you can predict the event or the date but not both. I am not sure you can predict either in the case of the use of nonparametric IV in applied research.

**Collaborations with Enno Mammen**

*Let's talk about your work with Enno Mammen.*

Enno and I met when he was at Humboldt University in Berlin. The topic of nonparametric additive models had arisen in statistics roughly 20 years earlier. Charles Stone had shown that if a conditional mean function of a multi-dimensional covariate is the sum of functions of a single covariate, then the individual functions can be estimated with the one-dimensional nonparametric rate of convergence. Additivity restricts the class of functions assumed to contain the conditional mean function and reduces its size relative to the size of the class used in fully nonparametric estimation without the additivity assumption. For example, if you make a series expansion of the conditional mean function, additivity greatly reduces the number of Fourier coefficients that must be estimated. Each coefficient can be estimated at a root-n rate, but the rate for the (truncated) series is an decreasing function of the total number of coefficients. If you have, say, a four-dimensional covariate and 10 series terms for each dimension, there are 10 to the fourth coefficients to estimate in fully nonparametric estimation. However, with an additive model, there are only 40 coefficients. Oliver Linton and several co-authors proposed a method called marginal integration, which integrates out all but one of the nonparametric additive components, but that method has a curse of dimensionality. It requires increasingly strong smoothness restrictions on the class of functions being estimated as the dimension of the covariate increases. Enno and I started thinking about whether there is a way to overcome this problem. We considered a nonparametric additive model with a link function and wondered if there is a way to estimate each additive component at the one-dimensional nonparametric rate. We eventually found a way to do this and achieve oracle efficiency. Oracle efficiency means that the estimate of each additive component has the same asymptotic distribution that it would have if all the other additive components were known. We were pretty happy about that and wrote a paper that was published in the *Annals of Statistics* [37]. We also considered the more difficult problem of estimating a nonparametric additive model with an unknown link function. We eventually found an estimation method, but it was less elegant than the estimator with a known link function and harder to compute. Our paper about this estimator was published in *Econometric Theory* [35].

Enno was at the University of Mannheim at that time. There was a visitor to the department named Jussi Klemelä. Enno, Jussi, and I wrote a paper about nonparametric additive models. It was much more of a mathematics paper and got published in *Bernoulli* [38]. Enno and I then considered estimation in a general class of functions based on a finite number of one-dimensional compositions, where each composition has a nonparametric additive structure. This amounts to estimating functions with scalar arguments, but they have a nested structure that makes estimating them much more complicated. We developed a rate-optimal least squares estimator of the additive components of a conditional mean function and showed how to extend it to conditional quantiles and neural networks. Our paper was published in the *Annals of Statistics* [36]. More recently, compositions of functions that are more general than ours and whose nesting structure is unknown have become important in the theory of neural networks. One example is a paper that Jian Huang (a statistician with whom I have written several papers), several mathematicians and computer scientists, and I have about using a neural network in which the data generation process has an unknown nesting structure to estimate a nonparametric quantile regression model with non-crossing estimated quantiles. That paper is forthcoming in the *Journal of Machine Learning Research* [33].

**Collaborations with Jian Huang**

*Before this recent paper with Jian Huang, you wrote with him several papers on sparse high dimensional regression models, right?*

Yes. Jian was in the statistics department at the University of Iowa. I attended a seminar he gave about estimating a high-dimensional nonlinear model. After it was over, we had a conversation about penalized estimation with a bridge penalty function. In a linear model, the LASSO penalty function is the sum of the absolute values of the slope coefficients. The bridge penalty function takes the absolute values to a power less than one. In contrast to the LASSO penalty function, the bridge penalty function is a concave function of the absolute values, which enables it to penalize coefficients that are close to zero more heavily than larger coefficients. This penalty function had been investigated in a paper by Keith Knight and Wenjiang Fu that was published in the *Annals of Statistics* in 2000. Jian and I found conditions under which a bridge estimator of the coefficients of a sparse linear model is model selection consistent and oracle efficient. That is, the estimator selects the variables with non-zero coefficients correctly with probability approaching 1 as the sample size increases, and the asymptotic distribution of the estimated non-zero coefficients is the same as that of the ordinary least squares estimator applied to the correct (but unknown) model. The LASSO, by contrast, is model selection consistent only under a restrictive condition called the irrepresentable condition, and LASSO estimates of the non-zero coefficients have an asymptotic bias due to the penalization. Computation of bridge estimates is hard, because the bridge objective function is non-convex. Shuangge Ma, a biostatistician at Yale whom Jian knew but I have never met, agreed to do the computation. In the end, the three of us wrote a paper that was published in the *Annals of Statistics* [46].

*Nowadays, LASSO and all this high dimensional stuff is extremely popular in econometrics or economics overall. But at the time I guess this was very remote within the community of econometrics. Can you now describe your other high-dimensional papers with Jian?*

We moved on to high-dimensional nonparametric additive models. Suppose you have a nonparametric additive structure, the first component being a function of the first regressor only, and so forth, but the number of nonparametric components may be very large. We proposed approximating each additive component with a truncated series expansion and using a group LASSO penalty function. This penalty function penalizes groups of coefficients so that the estimate of an additive component can be zero. We proved that the adaptive group LASSO estimate of a high dimensional nonparametric additive model is model selection consistent and oracle efficient. That led to another paper in the *Annals of Statistics* with a third co-author, Fengrong Wei, who had been Jian's student [47].

*For your work with Peter Hall, Enno Mammen, and Jian Huang, in the later stages of your career, it seems the Annals became quite a regular outlet for your work. In general, this journal is often regarded as being highly mathematical. Looking at your career, it appears to have taken an intriguing trajectory. Initially, you were more focused on applied econometric work, but later on, you shifted towards more theoretical aspects.*

People have commented on this. My career proceeded backwards!

*I wouldn't say it was backwards, but it's certainly evolved into a more mathematically intensive direction.*

I think of myself as working in the boundary area between econometric theory and mathematical statistics. There is a separation between what I'm interested in and what is fashionable in econometric theory these days. I have kept working on problems that are arguably closer to statistics. Some people say I am a crypto statistician, but I have reached a stage in life where I can work on what I want without jeopardizing my career.

**Recent Research Interests: High-Dimensional Models**

*Can you talk about your recent research on high-dimensional models?*

Cun-Hui Zhang and Jian Huang had written a paper published in the *Annals* about estimating a linear regression model under what they called a generalized sparsity condition. This condition allows the model to have some coefficients that are close but not equal to zero. These coefficients are smaller than random sampling errors, so they cannot be distinguished empirically from zero. Zhang's and Huang's paper is about LASSO estimation of such a model. Jian and I considered an extension. Suppose you have a linear model and the object of interest is a particular "large" coefficient, $\beta_1$ say.

Or suppose you have a nonparametric additive model, and the object of interest is a particular "large" additive component. The other coefficients or additive components may be large, zero, or close but not equal to zero. Dropping some or all of the non-zero but small coefficients or additive components produces an omitted variables bias but it also reduces the variance of the estimate of $\beta_1$ or the additive component of interest. If the reduction in variance is sufficiently large or the omitted variables bias is sufficiently small, the mean square estimation error of $\beta_1$ or the additive component of interest decreases. This is related to issues in applications fields such as labor economics. Suppose you have a loglinear wage equation in which log(wage) is equal to an intercept plus $\beta_1$ times years of education plus other variables. For purposes of this discussion, let's assume that all the variables are exogenous and that the loglinear model is correctly specified. Standard data sets for estimating wage equations contain hundreds of variables that are arguably related to the wage in some way. Speaking an obscure foreign language is an example. The influence of such a variable on the conditional mean wage is probably very small, but it is not zero; some people make a living by translating this language. However, dropping such variables from the model may reduce the mean square error of the estimate of $\beta_1$ because the variance of the estimate decreases and the omitted variables bias may be small. But there are many variables. Which, if any, should be left out of the wage equation to minimize the mean square error of the estimate of $\beta_1$? There are certain variables that everybody would agree should be included, but others are questionable. What happens if you apply a model selection consistent penalization method like the adaptive LASSO or SCAD to such a model? It won't be model selection consistent anymore, because model selection consistency means getting all the zeros and non-zeros, right, which is impossible if some of the non-zeros are very small. But you might be able to get a smaller mean square error of the estimate of the coefficient or function of interest. Jian and I wrote a paper giving conditions under which, with probability approaching one, several penalized least squares procedures distinguish correctly between large and small coefficients or additive components, drive the small ones to zero, and reduce the asymptotic mean square errors of the large coefficients or additive components. This paper was published in *Statistica Sinica* [48].

*Can you do some inference after this model selection, or is that a separate issue?*

Yes, if there are coefficients, (or parameters in a nonlinear model) that are sufficiently far from zero and others that are close to zero, but there is a little gap between them. For example, the large parameters might be larger than $O(n^{-1/2})$ and the small ones $o(n^{-1/2})$. Then you can do inference on the large parameters. You cannot do inference on the small ones.

*Can you use the bootstrap to get asymptotic refinements?*

If the total number of parameters is *n* to a power less than one, the number of large parameters is fixed, and all the other parameters are zero, then the bootstrap can provide asymptotic refinements for a model that may be nonlinear. These refinements are of the same order that would be achieved if the oracle or true model were known. If there are parameters that are close but not equal to zero, the bootstrap is consistent. It provides correct asymptotic inference but not necessarily refinements. In addition, Soumendra Lahiri and co-authors have given conditions under which the residual and perturbation bootstraps provide asymptotic refinements for a linear model in which the number of coefficients can exceed the sample size.

*In econometric applications, I think it's quite reasonable that p is smaller than n but p is growing. Do you have other high-dimensional papers?*

There is one with Lars Nesheim. He had a large data set from supermarkets in the U.K. about peoples' purchases of butter and margarine. In marketing, random coefficients models are often used to model choices among different products. We used a random coefficients logit model to choose among brands of butter and margarine. There were many coefficients in the model, and they were not necessarily all random. We used penalized maximum likelihood to select the variables that should be in the model and also determine which coefficients were random or not. The penalization method is adaptive LASSO but applied to maximum likelihood estimation instead of least squares. In addition to selecting variables and random vs fixed coefficients, we computed measures that are relevant to substantive economics such as consumer surplus and changes in market shares. Our paper was published in the *Journal of Econometrics* [19].

**Recent Research Interests: Shape Restrictions**

*Shall we now talk about your work on shape restrictions?*

The shape work grew out a search of the applied IV literature that my RA and I had done for the Fisher-Schultz paper. The instrument and endogenous explanatory variable were discrete in most of the applications of IV that we found. Moreover, the discrete endogenous variable often had more points of support than the discrete instrument. The structural function is neither point nor partially identified nonparametrically when the endogenous variable has more points of support than the instrument. Joachim Freyberger was a PhD student at Northwestern at the time. He and I investigated what happens if a shape restriction such as monotonicity or convexity is imposed on the structural function. These shape restrictions can be represented as linear inequalities. It turns out that under such shape restrictions, linear functionals of the structural function are partially identified. Identified lower and upper bounds on a linear functional are optimal values of the objective functions of linear programming problems. Linear programming has been understood for over 70 years and efficient algorithms for solving linear programming problems are well known. Moreover, the feasible region of a linear program is convex, which implies that the identified set of the linear functional is the entire interval between its lower and upper bounds. There are no holes in the interval consisting of values that the functional cannot take. The objective function of the linear program and the shape constraints do not depend on unknown population parameters, but there are also equality constraints that impose the moment conditions of IV estimation. The moment conditions and, therefore, the equality constraints depend on certain expected values. These are unknown population quantities, but they can be estimated consistently. Replacing the population expected values with consistent estimates gives consistent estimates of the bounds on the linear functional. The estimated bounds have random sampling error, so it is necessary to find a way to do inference on the bounds. This is complicated because the feasible region of a linear programming problem can have multiple optimal or near optimal vertices. The asymptotic distribution of the estimated bounds is the maximum or minimum of correlated multivariate normals when there are multiple optimal vertices. Near optimal vertices create a uniformity problem that can cause asymptotic approximations to be inaccurate in finite samples. Joachim and I found a bootstrap method for doing inference. In the setting we treat, the bootstrap is consistent, regardless of whether there are multiple optimal or near optimal vertices. In general, the bootstrap does not estimate the distribution of the maximum or minimum of several random variables consistently because the empirical distributions of the variables are not centered properly. In our setting, however, it is possible to center the empirical distributions correctly. We wrote a paper that was published in the *Journal of Econometrics* [8].

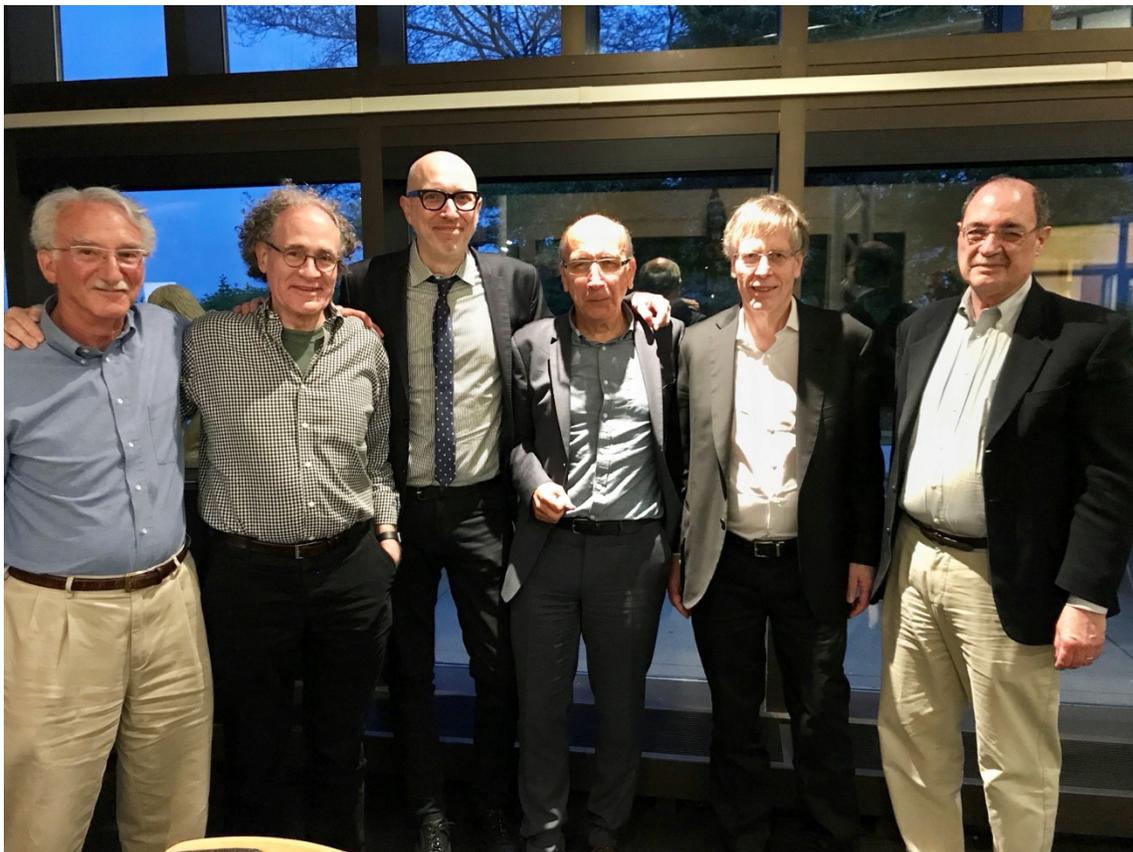

(Photo caption: Joel Horowitz, Larry Christiano, Adrian Randolph (Dean of the Weinberg College of Arts and Sciences at Northwestern), Richard Blundell, Lars Hansen, Charles Manski (from left to right), Dinner at which Richard Blundell received the Nemmers Prize in economics, May 2017)

*So that was your first shape-constraint paper. You also have some other shape-constraint papers with Richard Blundell.*

With Richard and his former student Matthias Parey. We estimated a demand function for gasoline nonparametrically using data on individuals' purchases. The explanatory variables were price and income. The estimated demand function turned out to be wiggly and non-monotonic in the price. This was due to random sampling error and, among other undesirable effects, caused estimates of welfare measures such as deadweight losses to have incorrect signs.  Monotonicity can be imposed on the estimated function in several ways, such as imposing a monotonicity constraint and rearrangement. However, we wanted a constraint that is consistent with the theory of the consumer. We used the Slutsky condition, which is a nonlinear inequality constraint. The estimated demand function is monotonic in the price when this constraint is imposed.  Moreover, we found that low- and high-income individuals are more sensitive to price changes than middle income individuals are, which is something not revealed by standard parametric models but clear in the nonparametric estimate.  We wrote a paper about this work that was published in *Quantitative Economics* [4].

In subsequent work, we let the demand function be non-separable in the unobserved heterogeneity variable. This led to a nonparametric quantile regression model with the Slutsky constraint. Our paper about that was published in the *Review of Economics and Statistics* (RESTAT) [3].  We also dealt with the problem that reported prices are regional averages, not the prices that people really pay. This creates a type of errors-in-variables problem called Berkson errors. In Berkson errors, the price paid is the average price plus a random variable.  This is the opposite of classical errors in variables. Peter Hall and Aurore Delaigle had published a paper in the *Journal of the Royal Statistical Society, Series B* showing that a function whose argument has Berkson errors is not nonparametrically identified unless there is exogenous information about the error distribution. There was also work by Susanne Schennach and others giving conditions under which a suitable instrument provides the required information and can be used to achieve nonparametric identification, but those conditions were not applicable to our setting.  Matthias found data that can be used to estimate the distribution of the Berkson errors and enabled us to identify the demand function. We then found a way to estimate a non-separable demand function with Berkson errors nonparametrically subject to the Slutsky constraint.  It turned out that Berkson errors had a substantial effect on the estimated demand function without the Slutsky constraint but a smaller effect with the constraint.  We also found a way to treat the possibility that the price is endogenous. This paper was also published in RESTAT [2]. In a separate paper published in the *Econometrics Journal*, Matthias, my former student Jia-Young Michael Fu (Mike), and I described a test for exogeneity in a nonparametric quantile model [9]. We used that test in the papers about non-separable demand functions with and without Berkson errors.

*Did your work on testing exogeneity in a quantile IV model grow out of your earlier work with Richard on testing exogeneity?*

Yes. Richard and I had developed a test of exogeneity in a nonparametric IV model in which identification is achieved through a conditional mean restriction. In a quantile IV model, identification is achieved through a conditional quantile restriction. The non-smoothness of quantile estimators makes testing exogeneity in a quantile model more difficult than it is in a model that is identified by a conditional mean restriction. Mike had been a graduate student at Northwestern. He started working on the quantile IV problem and made enough progress to give a talk about it at a conference for graduate students at the UCL economics department. However, he dropped out of the Northwestern graduate program for family reasons. I agreed to finish the work on the exogeneity test but kept delaying until the quantile testing problem arose in the work with Richard and Matthias. That motivated me to finish the work that Mike had started and write a paper. Matthias did the computational work. The paper was published in the *Econometrics Journal* some 10 years after Mike had started it [9]. He had been an undergraduate at Cornell and was friends with Francesca Molinari. She told me how to get in touch with Mike to send him the published paper.

*Was he happy?*

Yes. He was happy to learn that the work had been published. Of course, he is one of coauthors whose name is on the paper.

*Are you continuing the work with Richard and Matthias?*

Yes. A problem with the Berkson error model is that each individual is offered a price that is the regional average price plus the Berkson error, which is random. But that is not necessarily how things work in the real world. People may see a distribution of prices and choose from this distribution. Variables other than price affect the choice, so people do not necessarily choose the lowest price. We were aware of this problem when we did the Berkson work. but we could not do anything about it because our data did not tell us the prices people were offered or chose. Matthias is German and obtained a German data set that includes offered and chosen prices among many other variables. We have found that a nested logit model fits the data well. The first decision in the nest is whether to make a purchase, and the second decision is the choice of price conditional on making a purchase. The nested logit model makes predictions that are very different in substantively important ways from the Berkson error model. However, this is ongoing research, so we don't yet have a paper or firm results to report.

*Maybe this is potentially off the record, but can you comment a bit about the current state of econometric theory or generally econometrics?*

I think that until recently, econometric theory had diverged from substantive economics in many ways. Maybe my work is an example of that. I think econometrics has suffered from this divergence. But recently, things have happened that have brought econometric theory closer to substantive economics. One is causal inference, which of course is not just an issue in economics but is generating a lot of theoretical work in econometrics. Other areas that have brought econometrics and substantive economics closer together are the design of field experiments in, for example, development economics and the analysis of networks. Both present challenging identification and inference problems that have attracted the attention of econometric theorists.

**PUBLICATIONS OF PROFESSOR JOEL L. HOROWITZ CITED IN THE INTERVIEW**